# Comparative Computational Study of the Energetics of Li, Na, and Mg Storage in Amorphous and Crystalline Silicon

Fleur Legrain[1], Oleksandr I. Malyi[1], and Sergei Manzhos[1]

[1]Department of Mechanical Engineering, National University of Singapore, Block EA #07-08, 9 Engineering Drive 1, 117576, Singapore

Phone: *+65 6516 4605* E-mail: mpemanzh@nus.edu.sg (S. M.)

**ABSTRACT**

To assess the potential of amorphous Si (*a*-Si) as an anode for Li, Na, and Mg-ion batteries, the energetics of Li, Na, and Mg atoms in *a*-Si are computed from first-principles and compared to those in crystalline Si (*c*-Si). It is shown that Si preamorphization increases the average anode voltage and reduces the volume expansion of the anode during the insertion of the metal atoms. Analysis of computed formation energies of Li, Na, and Mg defects in *a*-Si and *c*-Si suggests that the energetics of the single atoms into *a*-Si are thermodynamically more favorable. For instance, defect formation energies of Li, Na, and Mg defects in *a*-Si are respectively 0.71, 1.72, and 1.82 eV lower compared to those in *c*-Si. Moreover, the defect formation energies of Li, Na, and Mg defects (vs. vacuum reference states) in *a*-Si are comparable with the metal cohesive energies and consequently the insertion of the metal atoms might be possible with appropriate control of charging process. This is in contrast to *c*-Si, where the storage of Na and Mg atoms is limited due to high energy cost of Na and Mg insertion into *c*-Si.

# 1. Introduction

The development of high energy density and/or high-rate electrochemical batteries is the key to sustainable development as they will enable large-scale storage of electricity derived from intermittent sources (such as wind and solar). The size and geographic distribution of Li resources indicate that Li-ion batteries alone, which are the most widely used metal-ion batteries today, cannot satisfy the increased needs in energy storages [1, 2]. Therefore, the interest in non-Li-ion batteries increases. Among the alternative technologies, Na and Mg-ion batteries are considered as the most attractive. It is known that these batteries have worse performance compared to those of Li-ion batteries [3-5]. Nevertheless, since Na and Mg resources are abundant, the use of both Na and Mg-ion batteries is attractive from the economic point of view.

Recent investigations of cathode materials for both Na [6-8] and Mg-ion [9-11] batteries reported promising results. This puts the onus on the development of other battery components including electrolytes and negative electrodes, to achieve commercialization. Specifically, the design of anode materials for these batteries is a big challenge. Similar to metallic Li anodes, the use of metallic Na [1] and Mg [5, 12] is disadvantageous because of dendrite formation and undesired reactions with electrolyte species (especially for Mg-ion batteries), leading to capacity fading as well as safety problems. Replacing metallic Mg with an insertion/intercalation type anode materials would allow a degree of control over the insertion voltage that could mitigate this problem. Hence, the development of insertion/intercalation type anode materials for both Na and Mg-ion batteries is important. Unfortunately, anode materials which are attractive for Li-ion batteries are often not attractive for Na and Mg-ion batteries. As illustration, it was reported that Na and Mg does not intercalate into graphite [13-15] and crystalline Si ($c$-Si) [16, 17]. The difference in the behavior of Li and Na/Mg atoms comes from both sizes of the metal atoms and their electronegativities. Hence, the appropriate design of anode materials requires detailed theoretical and experimental studies.

Recently, it has been shown that Sn can be used as attractive anode material for both Na [17-19] and Mg-ion [20, 21] batteries. However, Sn is significantly heavier compared to Si, hence, the theoretical specific capacities of Sn-based materials are smaller compared to that of Si [3, 21]. Several experimental studies showed that preamorphization of an anode material can significantly improve (make the insertion of metal atoms more thermodynamically favorable) the energetics of inserted atoms [19, 22], and in some cases, it could enable reversible insertion/deinsertion of metal atoms. For instance, it has been shown experimentally that amorphous phosphorus [22] materials can be used as anode materials for Na-ion batteries. In contrast, the insertion process into the bulk crystalline materials is limited. Recently, Kaxiras's group reported first-principles investigation of the behavior of Li atoms in $c$-Si and amorphous Si ($a$-Si), and they found that the preamorphization of $c$-Si leads to improved insertion energetics [23]. How does Si preamorphization change the energetics of inserted non-Li metal atoms? For our best knowledge, this question is still open. Therefore, in this paper, we report comparative computational studies of the energetics of Li, Na, and Mg atoms in $c$-Si and $a$-Si structures to explain how Si preamorphization changes the performance of anode materials for different metal-ion batteries.

## 2. Methods

All calculations were carried out using density functional theory (DFT) [24] and the SIESTA code [25]. The Perdew-Burke-Ernzerhof (PBE) exchange-correlation functional [26] and the double-polarized orbitals (DZP) basis set were used. We used a standard basis set as generated by SIESTA, but the cut-off radii were slightly increased from the default values by choosing $E_{shift} = 0.01$ Ry. A cutoff of 100 Ry was used for the Fourier expansion of the density, and Brillouin-zone integrations were done with a 3×3×3 $k$-point Monkhorst-Pack mesh [27]. Core electrons were treated within the effective core approximation with Troullier-Martins [28] pseudopotentials (provided with SIESTA, see supporting information). Geometries were optimized until forces on all atoms were below 0.02 eV/Å.

The amorphous silicon (*a*-Si) was modeled by a 64-atom simulation cell with periodic boundary conditions. The input structure was taken from Ref. [29] which used a similar computational setup. To predict Li, Na, and Mg-inserted structures, we used insertion sites reported in Ref. [29] as initial guesses. The crystalline silicon (*c*-Si) was modeled by a 2×2×2 supercell (64-atoms). We considered both vacuum (modeled as a cubic cell of size 11×11×11 Å$^3$) and bulk (body-centered cubic cells for both Li and Na and hexagonal cell for Mg) reference states for Li, Na, and Mg. All structures were fully relaxed until the stresses were below 0.1 GPa. We performed spin-polarized calculations for most considered structures, however, in most cases, it was found insignificant.

## 3. Results and discussion
### 3.1. Voltages and volumetric energy densities

*a*-Si is metastable structure under normal conditions. Computed results indicate that the cohesive energy of *a*-Si is 0.13 eV per atom larger (*a*-Si is less stable) than that of *c*-Si. This observation indicates that an *a*-Si anode will have a larger average anode voltages compared to those of *c*-Si. For instance, if we assume that the charging of the anode material goes according to eq. 1

$$Si + xM \leftrightarrow SiM_x \tag{1}$$

Then, using the well-defined methodology (see supporting information) [4, 30], the average anode voltages ($V$) vs. metallic reference states can be calculated according to eq. 2

$$V = -\frac{E_{M_xSi} - E_{Si} - xE_M}{zxe} \tag{2}$$

where $E$ is the corresponding DFT energy; $z$ and $e$ are the charge (in electrons) and absolute value of the electron charge. Assuming that fully charged Si anodes can be reached (in this work, we considered that the final charge states are $Li_{3.75}Si$ ($Li_{15}Si_4$), $Mg_2Si$, and $NaSi$) and taking into account the number of valence electrons for the considered metal atoms (1 for Li and Na, and 2 for Mg), we estimated the average voltage of *c*-Si and *a*-Si anodes (see Table 1). The predicted voltages for *c*-Si-based alloys are comparable to the previously reported data but show small underestimations. For instance, the predicted voltage for Mg-ion battery is 0.06 V smaller than the reported value (0.15 V) [21]. The difference can be caused by the difference in computational methods, e.g. between an atomic-centered basis used here and plane-wave calculations reported previously [25, 31]. On the one hand, the calculations predict that a metal-ion battery with an *a*-Si anode will have a lower average battery voltage (higher average anode voltage) than the same battery with a *c*-Si anode. The difference of average battery voltages ($\Delta V$) can be calculated according to eq. 3 (see supporting information):

$$\Delta V = \frac{\Delta E_{coh}}{zxe} \tag{3}$$

where $\Delta E_{coh}$ is the difference of cohesive energies of *c*-Si and *a*-Si. Eq. 3 suggests that the largest voltage difference is expected for Na-ion batteries ($zx=1$), while for the Mg ($zx=4$) and Li-ion ($zx=3.75$) batteries the differences are significantly smaller (see Table 1). On the other hand, since it is well known that many electrolytes are not stable within the battery operating voltage [32], the larger anode voltage (lower battery voltage) may palliate the electrolyte stability problem and improve cycle life and safety of the battery. Although the anode voltage is an important parameter, it is also critical to understand the performance of anode materials in light of volume expansion, as insertion type anodes, and specifically Si, are notorious for large volume changes upon cycling that creates hefty design problems [33, 34]. As a criterion, we will use the volumetric energy density as proposed by Obrovac and co-workers (see supporting information) [34]. Namely, the volumetric energy density ($\tilde{U}_f$) of a charged negative electrode alloy material can be calculated according to eq. 4:

$$\tilde{U}_f = \frac{FV_{avg}}{v}\left(\frac{\xi_f}{1+\xi_f}\right) \qquad (4)$$

where $F$ is Faraday's number in units of Ah/mol ($F \approx 26.802$ Ah/mol); $\xi_f$ is the percent volume expansion; $V_{avg}$ is the average voltage of the full cell (in this work, we used 3.75 V as the cathode voltage); $v$ is the volume occupied by metal atom per valence charge. Computed maximum volumetric energy densities suggest that at the fully charged states batteries with $c$-Si anodes will have a slightly larger volumetric energy densities than the same batteries with $a$-Si anode (compare where curves for $a$-Si and $c$-Si end in Fig. 1). However, different volume occupied by a Si atom in $c$-Si (~20.82 Å$^3$) and in $a$-Si (~21.40 Å$^3$) implies that maximum volume expansions of $c$-Si anodes are slightly larger than those of a-Si. For instance, for Mg-ion batteries the volume expansion of $c$-Si (220 %) is 7% larger than that of $a$-Si (see Table 1). This implies that the difference of volumetric energy densities of the anode materials at the same volume expansion is different with that of maximum volumetric energy densities. As an illustration, as it can be seen from the Fig. 1, at the same volume expansion (which may be limited by design considerations), the differences between the volumetric energy densities of $a$-Si and $c$-Si anodes are insignificant. Therefore, at the appropriate design of anode materials, the higher voltage of an $a$-Si anode (lower battery voltage) should not result in a lower volumetric energy density while being advantageous for battery stability.

**3.2. Defect formation energies**

The defect formation energy ($E_f$) was calculated according to eq. 5.

$$E_f = E_{Si+M} - E_{Si} - \mu_M, \qquad (5)$$

where $E_{Si+M}$ and $E_{Si}$ are the DFT energies of doped and undoped silicon systems; $\mu_M$ is the chemical potential of metal atom (Li, Na, or Mg). It is well known that the chemical potential differs between cathode, electrolyte, and anode. The difference between chemical potentials of Li in anode and cathode determines the battery voltage [30]. To analyze the insertion of metal atoms into anode materials, recent studies calculated $\mu_M$ as energy per metal atom in a vacuum [16, 21, 35-40] or in bulk metal [41, 42]. Defect formation energy calculated vs. vacuum or metal reference state is certainly useful. For instance, the negative defect formation energies with respect to the vacuum reference states suggest that metal atoms can be inserted into the electrode, if there is no metal clustering at the material surface. The defect formation energies with respect to the bulk metal reference states are also of significant interest. If we assume that the metal atom does not change the energetics of sub-surface layers or this change is small (usually, this is the case for materials whose structures are preserved during the insertion of metal atoms. *c-Si and a-Si are such materials only for extremely low dopant concentrations [41, 43, 44]*) and neglect the effects of the metal surface, the positive defect formation energy with respect to the bulk metal reference state suggests that for the metal atom, it is thermodynamically favorable to be inside of a large metal cluster than to diffuse into the electrode material. In contrast, negative defect formation energy suggests that formation of metal clusters is thermodynamically unstable.

Computed defect formation energies of Li, Na, and Mg defects in *c*-Si suggest that metal atoms occupy interstitial tetrahedral (Td) sites (see **Fig.** 2a and supporting information). Despite this similarity of defective structures among the three metals, the defect formation energies are significantly different. For Li atoms, the defect formation energy vs. vacuum reference state is negative suggesting that the insertion of single metal atoms is thermodynamically favored vs. vacuum reference state. However, due to a low cohesive energy (-1.70 eV), the defect formation energy vs. metal reference state is positive (0.52 eV). This indicates that there is a cost for the insertion of the single atom vs. metallic reference states. *Nevertheless, it does not imply that the insertion cannot happen.* First, a Li atom adsorbed on a Si surface has a positive charge [16] and a small overpotential can lead to the insertion of the metal atom. Moreover, it has been shown that the insertion of metal atoms into *c*-Si weakens the Si-Si [40, 41, 44, 45] bonds leading to improved insertion. The predicted defect formation energies for Na and Mg defects in *c*-Si are 1.82 and 2.52 eV vs. metallic reference states (see Table 2). Because of this and due to slow dopant diffusion [16, 21, 40], it is expected that the insertion of these metal atoms into *c*-Si is limited at realistic charge overpotentials and rates, which is consistent with recent theoretical and experimental predictions [16].

For *a*-Si, the defect formation energies calculated for lowest energy configurations (see Fig. 2b-2c) are significantly smaller than the corresponding values for *c*-Si. For instance, the lowest defect formation energy of Li defect is 0.71 eV smaller than that in *c*-Si (see Table 2). This is consistent with the recent observation from the Kaxiras's group which reported the difference of 0.79 eV [23]. The predicted defect formation energies indicate that the insertion of Li atoms into *a*-Si is thermodynamically favorable vs. both reference states. Hence, Si preamorphization leads to significantly improved energetics of Li storage. For both Na and Mg atoms, a significant improvement of the insertion energetics is also found (see Table 2). For instance, the Na and Mg defect formation energies vs. vacuum reference states are negative. The defect formation energies vs. metallic reference states are positive (0.10 eV and 0.69 eV for Na and Mg, respectively). The predicted defect formation energies are significantly lower than those in *c*-Si. Moreover, the computed defect formation energies are comparable with those for Li defects in c-Si. Since, similar to the insertion of metal atoms into c-Si, the insertion of metal atoms in *a*-Si leads to the charge transfer from metal atoms to Si matrix, it is expected that defect formation can be improved by overpotential and/or formation of defect clusters (similar to *c*-Si [41]). Obviously, the overpotential for the insertion of single Mg atom is expected to be the largest among the three considered doped amorphous systems. Nevertheless, taking into account the presence of two valence electrons per Mg atom, we can conclude that it is comparable with that for Li insertion into ideal *c*-Si (see supporting information). Considering the charging process of anode materials (possibility to control the charge/discharge voltage, use of a charge overpotential, negative formation heats of SiNa (-0.11 eV) and SiMg$_2$ (-0.48 eV) vs. *a*-Si and metallic reference states, negative charges of metal atoms absorbed/inserted into Si [16, 40, 46], the negative defect formation energies vs. vacuum reference states (see Table 2), and possibility of defect cluster formation [41]), we believe that appropriate control of charge/discharge process may achieve the insertion of these metal atoms into *a*-Si.

**Conclusions**

In conclusion, using first principle calculations, we have studied the energetics of Li, Na, and Mg atoms in *a*-Si and *c*-Si. We found that Si preamorphization increases the average anode voltage (by up to 0.13 eV for Na) and reduces the anode volume expansion (by up to 9% for $Li_{3.75}Si$). Despite this, at the same volume expansion of the anode material, metal-ion batteries with *c*-Si and *a*-Si will have the same volumetric energy densities. We found that Si preamorphization reduces the defect formation energies of Li, Na, and Mg defects. For Li, this means that Li atoms will insert easier into *a*-Si (-0.18 eV vs. Li in bulk) compared to *c*-Si (0.52 eV vs. Li in bulk). Regarding Na, results show that single Na atom has a high insertion cost into *c*-Si (0.42 eV vs. Na in vacuum and 1.82 eV vs. Na in bulk) but it is much smaller for *a*-Si (-1.30 eV vs. Na in vacuum and 0.10 eV vs. Na in bulk). For Mg, Si amorphization also reduces the defect formation energies (2.52 eV in c-Si and 0.69 eV in a-Si, both vs. Mg in bulk). Taking into account the negative formation heats of NaSi (-0.11 eV) and $Mg_2Si$ (-0.48 eV) vs. metallic Mg and a-Si, we believe that appropriate control of charge/discharge process may realize the insertion of the metal atoms into *a*-Si. In the present work, we have thus demonstrated that *a*-Si was advantageous over *c*-Si for Li, Na, and Mg insertion energetics and volume expansions.

**Acknowledgments**

This work was supported by the Tier 1 AcRF Grant by the Ministry of Education of Singapore (R-265-000-430-133).

**Supporting information**

Information on methods and computed structures are available as supporting information.

**Table** 1. Average voltages (V) and relative volume expansions (%) of *c*-Si and *a*-Si anodes for Li, Na, Mg-ion batteries

|  | Li | | Na | | Mg | |
|---|---|---|---|---|---|---|
|  | $Si_4Li_{15}$ | | SiNa | | $SiMg_2$ | |
|  | Voltage | Expansion | Voltage | Expansion | Voltage | Expansion |
| *a*-Si | 0.27 | 292 | 0.11 | 128 | 0.12 | 212 |
| *c*-Si | 0.23 | 303 | -0.02 | 135 | 0.09 | 220 |

**Table** 2. Defect formation energies (in eV) of Li, Na, and Mg defects vs. bulk and vacuum reference states

|  | Li | | Mg | | Na | |
|---|---|---|---|---|---|---|
|  | *c*-Si | *a*-Si | *c*-Si | *a*-Si | *c*-Si | *a*-Si |
| vs. bulk | 0.52 | -0.18 | 2.52 | 0.69 | 1.82 | 0.10 |
| vs. vacuum | -1.27 | -1.98 | 0.92 | -0.90 | 0.42 | -1.30 |

**Figure captions**

**Figure** 1. Volumetric energy density for X-Si alloys as a function of volume expansion. Color and black lines represent results for c-Si and a-Si, respectively. Lines end at the maximum state of charge.

**Figure** 2. The lowest energy site for metal insertion in *c*-Si (a) and *a*-Si (b). Metal atoms are in blue color and Si host atoms in yellow. The difference in geometry among Li, Na, Mg insertion configurations is indistinguishable to the eye. Due to the similarity of different insertion sites, we show only one site which is the lowest energy site for Li and Mg and within 0.01 eV from the most equilibrium configuration of Na defects. (c) Computed defect formation energies of Li, Na, and Mg in *a*-Si vs. vacuum and bulk metal reference states. Corresponding values for *c*-Si are shown as black circles.

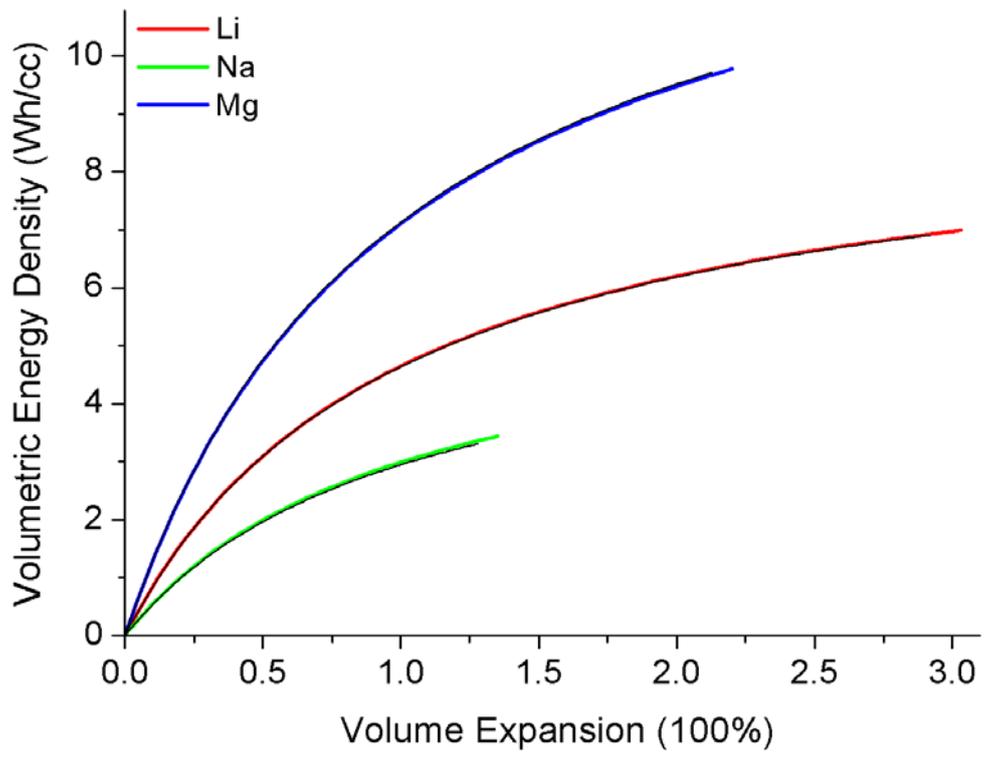

**Figure** 1

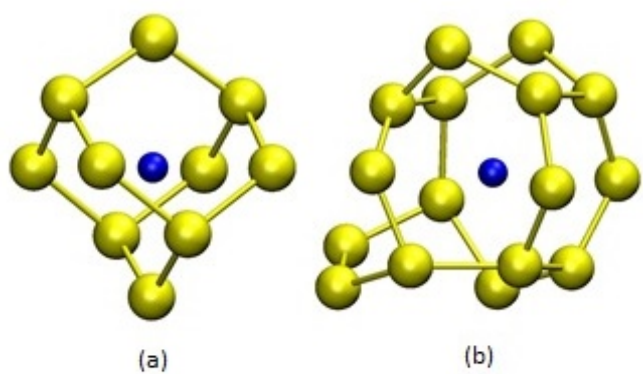

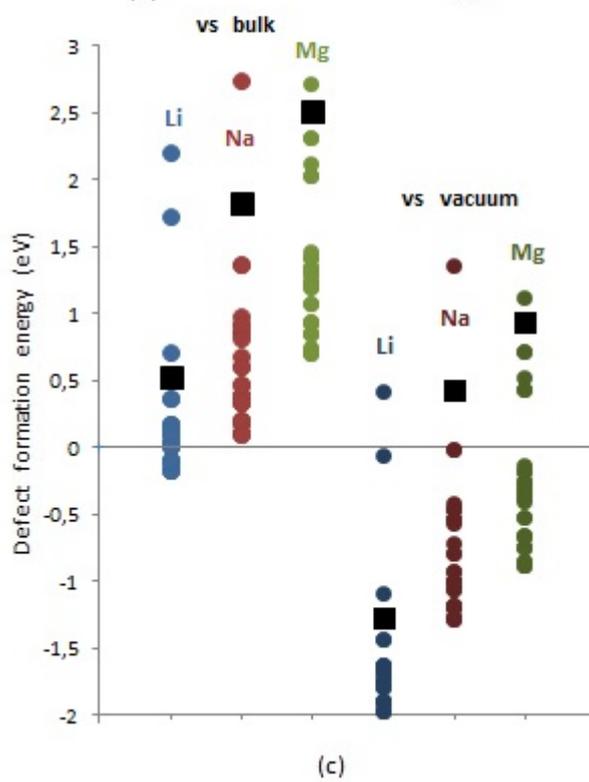

**Figure** 2

# SUPPORTING INFORMATION

for

# Comparative Computational Study of the Energetics of Li, Na, and Mg Storage in Amorphous and Crystalline Silicon

Fleur Legrain[1], Oleksandr I. Malyi[1], and Sergei Manzhos[1]

[1]Department of Mechanical Engineering, National University of Singapore, Block EA #07-08, 9 Engineering Drive 1, 117576, Singapore

This document includes additional information on the theory behind voltage and volumetric energy density calculations (sections A-D) as well as on the computational setup (section E). Sections F and G provide structural information of Si, doped Si, and metals.

## A. Electrochemical potential

The electrochemical potential ($\bar{\mu}$) combines the chemical and electrostatic potentials. It is expressed as follows:

$$\bar{\mu}_i = \mu_i + z_i F \phi$$

Where $\mu_i$ is the chemical potential of the species $i$, $z_i$ is the charge of the ion $i$, $F$ is Faraday's constant, $\phi$ is the local electrostatic potential.

## B. Volumetric energy density

In a metal-ion cell, the energy released during a time $dt$ is:

$$dE = [V_{(+)}(x) - V_{(-)}(x)] i dt$$

Where $x$ is the number of moles of metal per mole of silicon, $V_{(+)}$ is the cathode voltage, $V_{(-)}$ is the anode voltage, and $i$ is the electrical current.

The electrical charge transferred ($idt$) can also be written as:

$$idt = dq = zNedx = zFdx$$

Where $N$ is the Avogadro number, $e$ is the elementary charge, and $F$ is Faraday's number.

The volumetric energy density of an anode material at full charge can thus be expressed as:

$$\widetilde{U}_f = \frac{-\int_{x=x_f}^{0} [V_{(+)}(x) - V_{(-)}(x)] zF dx}{v(x_f)}$$

Where $x_f$ is the number of moles of metal per mole of silicon at full charge and $v$ is the molar volume of the anode.

The volumetric energy density can be simplified as follows:

$$\widetilde{U_f} = V_{AVG} F \frac{z x_f}{v(x_f)}$$

The molar volume of M (Li, Na, and Mg)-Si alloy has shown to be in a good approximation a linear function of the metal content [1, 2]:

$$v(x) = v_0 + z k_M x$$

Where $v_0$ is the molar volume of silicon and $k_M$ is volume occupied per unit charge stored in the $M_xSi$ alloy.

The substitution of the molar volume expression into the volumetric energy density gives:

$$\widetilde{U_f} = V_{AVG} F \frac{z x_f}{v_0 + z k_M x_f}$$

The volume expansion for the anode can be written as:

$$\xi = \frac{v(x) - v_0}{v_0}$$

By substituting the molar volume expression, the volume expansion can also be expressed as follows:

$$\xi = \frac{k_M x z}{v_0}$$

The volumetric energy density can thus be written in terms of the volume expansion [2]:

$$\widetilde{U_f} = V_{AVG} \frac{F}{k_M} \frac{\xi_f}{1 + \xi_f}$$

## C. Voltage

In order to obtain the volumetric energy density, the average anode voltage is needed:

$$V_{AVG}^{anode} = \frac{1}{x_f} \int_0^{x_f} V_M^{anode}(x) dx$$

$$V_{AVG}^{anode} = -\frac{1}{x_f} \int_0^{x_f} \frac{\mu_M^{anode}(x) - \mu_M^0}{zeN} dx$$

Where $x_f$ is the number of atoms of metal inserted in the anode per atom of silicon at full charge.

At $x = 0$, the anode consists in pure crystalline silicon $Si$ while at $x = x_f$ it consists in $M_{x_f}Si$, which allows us to write:

$$\int_0^{x_f}(\mu_M^{anode}(x) - \mu_M^0)dx = G_{M_{x_f}Si} - G_{Si} - x_f G_M$$

Where $G_{M_{x_f}Si}$, $G_{Si}$, and $G_M$ are respectively the Gibbs free energies of $M_{x_f}Si$, pure crystalline $Si$, and metal $M$ in bulk.

The average anode voltage can hence be expressed as follows [3]:

$$V_{AVG}^{anode} = -\frac{G_{M_{x_f}Si} - G_{Si} - x_f G_M}{x_f z e N} \approx -\frac{E_{M_{x_f}Si} - E_{Si} - x_f E_M}{x_f z e}$$

Where $E_{M_{x_f}Si}$, $E_{Si}$, and $E_M$ are the DFT energies of $M_{x_f}Si$, pure silicon, and pure metal, respectively.

## D. Voltage difference between a-Si and c-Si

As the final state of charge is the same, the voltage difference ($\Delta V$) between amorphous and crystalline silicon anodes can be expressed as follows:

$$\Delta V = V_{a-Si}^{anode} - V_{c-Si}^{anode} = -\frac{E_{M_{x_f}Si} - E_{a-Si} - x_f E_M}{x_f z e} + \frac{E_{M_{x_f}Si} - E_{c-Si} - x_f E_M}{x_f z e} = \frac{E_{a-Si} - E_{c-Si}}{x_f z e}$$

$$= \frac{E_{vac} + E_{a-Si}^{coh} - E_{vac} - E_{c-Si}^{coh}}{x_f z e} = \frac{\Delta E_{coh}}{x_f z e}$$

Where $E_{a-Si}$ and $E_{c-Si}$ are the energies for amorphous and crystalline silicon, respectively, and $\Delta E_{coh}$ is the difference of cohesive energies of amorphous and bulk Si.

## E. Methods

### E. 1. Pseudopotentials

We used Troullier-Martins pseudopotentials provided with SIESTA. As pseudopotentials distributed with SIESTA are occasionally updated, to insure reproducibility of our results, we provide below the headers of the pseudopotential files (psf format) that we used, including electronic structure and cutoff radii:

Si Si pb nrl pcec

 ATM3     24-JUN-12 Troullier-Martins
3s 2.00  r= 1.75/3p 2.00  r= 1.94/3d 0.00  r= 2.09/4f 0.00  r= 2.09/
  4  0 1074  0.177053726905E-03  0.125000000000E-01   4.00000000000

Li Li pb nrl pcec

 ATM3     24-JUN-12 Troullier-Martins
2s 1.00  r= 2.26/2p 0.00  r= 2.26/3d 0.00  r= 2.59/4f 0.00  r= 2.59/

```
 4  0  950  0.826250725555E-03  0.125000000000E-01   1.00000000000
```

Na Na pb nrl pcec

```
 ATM3     24-JUN-12 Troullier-Martins
3s 1.00  r= 2.83/3p 0.00  r= 2.83/3d 0.00  r= 3.13/4f 0.00  r= 3.13/
  4  0 1054  0.225341106970E-03  0.125000000000E-01   1.00000000000
```

Mg Mg pb nrl pcec

```
 ATM3     24-JUN-12 Troullier-Martins
3s 2.00  r= 2.18/3p 0.00  r= 2.56/3d 0.00  r= 2.56/4f 0.00  r= 2.56/
  4  0 1061  0.206562681389E-03  0.125000000000E-01   2.00000000000
```

E. 2. Basis set

We used a standard basis set as generated by SIESTA, but the cut-off radii were slightly increased from the default values by choosing the parameter $E_{\text{shift}} = 0.01$ Ry (resulting in broader basis functions vs. the default) to mitigate basis set superposition errors (BSSE). BSSE was estimated by (i) computing counterpoise correction and (ii) using a TZP (triple-$\zeta$) basis (with and without counterpoise correction), for several insertion sites. From these tests, BSSE is estimated to be of the order of 0.1 eV and is of a similar sign and magnitude for all insertion sites. Our conclusions are therefore not affected by BSSE.

A cutoff of 100 Ry was used for the Fourier expansion of the density. We confirmed that the results do not appreciably change compared to a 200 Ry cutoff.

A 3x3x3 $k$-point mesh was used. The results did not appreciably change when going from a 2x2x2 to a 3x3x3 $k$-point mesh

## F. Structures

F. 1. Amorphous Si structures

*F. 1. a) Pure amorphous Si structure*

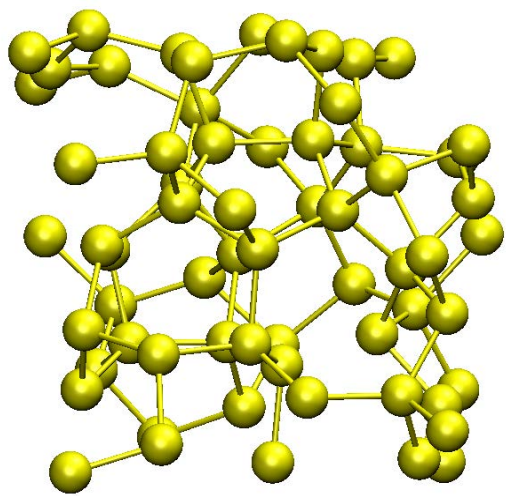

Figure 1 – The 64-atom simulation supercell of amorphous Si.

Lattice vectors (Å)

|   | a | b | c |
|---|---|---|---|
| a | 11.067 | -0.175 | 0.187 |
| b | -0.175 | 11.18 | -0.077 |
| c | 0.188 | -0.077 | 11.075 |

Fractional coordinates of the 64 atoms of Si

|    | a | b | C |
|----|---|---|---|
| Si | 0.354 | 0.920 | 0.793 |
| Si | 0.338 | 0.645 | 0.339 |
| Si | 0.289 | 0.720 | 0.848 |
| Si | 0.639 | 0.594 | 0.198 |
| Si | 0.459 | 0.491 | 0.264 |
| Si | 0.729 | 0.917 | 0.253 |
| Si | 0.539 | 0.249 | 0.513 |
| Si | 0.817 | 0.463 | 0.596 |
| Si | 0.537 | 0.723 | 0.069 |
| Si | 0.581 | 0.183 | 0.852 |
| Si | 0.735 | 0.727 | 0.344 |
| Si | 0.562 | 0.965 | 0.817 |
| Si | 0.276 | 0.086 | 0.106 |
| Si | 0.791 | 0.187 | 0.813 |
| Si | 0.823 | 0.917 | 0.058 |
| Si | 0.256 | 0.080 | 0.894 |
| Si | 0.433 | 0.596 | 0.936 |
| Si | 0.390 | 0.436 | 0.067 |
| Si | 0.672 | 0.829 | 0.939 |
| Si | 0.258 | 0.276 | 0.815 |
| Si | 0.338 | 0.933 | 0.578 |
| Si | 0.320 | 0.608 | 0.666 |
| Si | 0.137 | 0.984 | 0.535 |
| Si | 0.898 | 0.366 | 0.764 |
| Si | 0.887 | 0.161 | 0.487 |
| Si | 0.623 | 0.744 | 0.533 |
| Si | 0.054 | 0.019 | 0.894 |
| Si | 0.753 | 0.317 | 0.453 |
| Si | 0.548 | 0.287 | 0.041 |
| Si | 0.086 | 0.696 | 0.911 |
| Si | 0.001 | 0.938 | 0.703 |
| Si | 0.718 | 0.422 | 0.094 |
| Si | 0.195 | 0.374 | 0.008 |
| Si | 0.044 | 0.526 | 0.030 |
| Si | 0.464 | 0.298 | 0.721 |
| Si | 0.172 | 0.501 | 0.367 |
| Si | 0.035 | 0.860 | 0.032 |
| Si | 0.056 | 0.899 | 0.352 |
| Si | 0.523 | 0.972 | 0.283 |
| Si | 0.842 | 0.369 | 0.254 |

| Si | 0.830 | 0.046 | 0.654 |
| Si | 0.211 | 0.296 | 0.352 |
| Si | 0.101 | 0.329 | 0.683 |
| Si | 0.946 | 0.705 | 0.373 |
| Si | 0.896 | 0.035 | 0.318 |
| Si | 0.486 | 0.507 | 0.746 |
| Si | 0.513 | 0.036 | 0.493 |
| Si | 0.385 | 0.808 | 0.205 |
| Si | 0.668 | 0.591 | 0.670 |
| Si | 0.648 | 0.940 | 0.618 |
| Si | 0.425 | 0.298 | 0.346 |
| Si | 0.150 | 0.512 | 0.580 |
| Si | 0.143 | 0.235 | 0.161 |
| Si | 0.883 | 0.111 | 0.985 |
| Si | 0.192 | 0.901 | 0.176 |
| Si | 0.936 | 0.191 | 0.176 |
| Si | 0.998 | 0.533 | 0.250 |
| Si | 0.405 | 0.737 | 0.520 |
| Si | 0.970 | 0.616 | 0.567 |
| Si | 0.958 | 0.734 | 0.746 |
| Si | 0.464 | 0.153 | 0.189 |
| Si | 0.098 | 0.196 | 0.512 |
| Si | 0.772 | 0.670 | 0.837 |
| Si | 0.852 | 0.494 | 0.936 |

*F. 1. b) Lowest energy site structures*

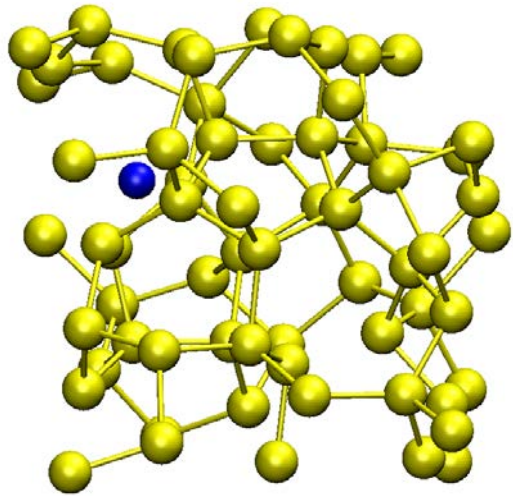

Figure 2 - Yellow and blue circles are Si and lowest energy sites for dopant atoms. Due to the similarity of different insertion sites, we show only one site which is the lowest energy site for Li and Mg and within 0.01 eV from the most equilibrium configuration of Na defects.

Li

Lattice vectors (Å)

|   | a      | b      | c      |
|---|--------|--------|--------|
| a | 11.104 | -0.178 | 0.129  |
| b | -0.179 | 11.159 | -0.071 |
| c | 0.130  | -0.072 | 11.103 |

Fractional coordinates of the 64 atoms of Si and 1 atom of Li

|    | a      | b     | c     |
|----|--------|-------|-------|
| Si | 0.354  | 0.923 | 0.795 |
| Si | 0.339  | 0.645 | 0.338 |
| Si | 0.289  | 0.722 | 0.847 |
| Si | 0.639  | 0.594 | 0.199 |
| Si | 0.460  | 0.491 | 0.263 |
| Si | 0.729  | 0.918 | 0.252 |
| Si | 0.552  | 0.251 | 0.514 |
| Si | 0.817  | 0.464 | 0.594 |
| Si | 0.537  | 0.723 | 0.068 |
| Si | 0.581  | 0.185 | 0.852 |
| Si | 0.736  | 0.728 | 0.344 |
| Si | 0.563  | 0.966 | 0.818 |
| Si | 0.276  | 0.087 | 0.106 |
| Si | 0.790  | 0.189 | 0.813 |
| Si | 0.824  | 0.918 | 0.057 |
| Si | 0.255  | 0.084 | 0.894 |
| Si | 0.433  | 0.595 | 0.935 |
| Si | 0.388  | 0.436 | 0.067 |
| Si | 0.672  | 0.829 | 0.939 |
| Si | 0.258  | 0.285 | 0.823 |
| Si | 0.339  | 0.930 | 0.579 |
| Si | 0.319  | 0.609 | 0.666 |
| Si | 0.135  | 0.984 | 0.537 |
| Si | 0.897  | 0.368 | 0.764 |
| Si | 0.884  | 0.161 | 0.485 |
| Si | 0.624  | 0.743 | 0.533 |
| Si | 0.054  | 0.021 | 0.894 |
| Si | 0.759  | 0.321 | 0.444 |
| Si | 0.547  | 0.288 | 0.041 |
| Si | 0.087  | 0.698 | 0.909 |
| Si | 0.000  | 0.940 | 0.704 |
| Si | 0.718  | 0.424 | 0.093 |
| Si | 0.190  | 0.376 | 0.015 |
| Si | 0.042  | 0.530 | 0.032 |
| Si | 0.460  | 0.296 | 0.720 |
| Si | 0.173  | 0.503 | 0.368 |
| Si | 0.036  | 0.861 | 0.031 |
| Si | 0.057  | 0.900 | 0.351 |
| Si | 0.524  | 0.973 | 0.282 |
| Si | 0.848  | 0.372 | 0.250 |
| Si | 0.829 | 0.048 | 0.655 |
| Si | 0.213 | 0.297 | 0.355 |
| Si | 0.101 | 0.331 | 0.687 |
| Si | 0.946 | 0.707 | 0.374 |
| Si | 0.896 | 0.036 | 0.316 |
| Si | 0.484 | 0.505 | 0.745 |
| Si | 0.516 | 0.037 | 0.492 |
| Si | 0.386 | 0.808 | 0.203 |
| Si | 0.666 | 0.589 | 0.669 |
| Si | 0.647 | 0.939 | 0.620 |
| Si | 0.429 | 0.298 | 0.349 |
| Si | 0.149 | 0.513 | 0.581 |
| Si | 0.144 | 0.235 | 0.165 |
| Si | 0.884 | 0.112 | 0.983 |
| Si | 0.193 | 0.902 | 0.176 |
| Si | 0.937 | 0.192 | 0.175 |
| Si | 1.000 | 0.537 | 0.249 |
| Si | 0.407 | 0.735 | 0.519 |
| Si | 0.969 | 0.617 | 0.567 |
| Si | 0.957 | 0.735 | 0.746 |
| Si | 0.464 | 0.154 | 0.190 |
| Si | 0.092 | 0.199 | 0.514 |
| Si | 0.771 | 0.671 | 0.835 |
| Li | 0.851 | 0.496 | 0.936 |

Na

Lattice vectors (Å)

| | | | |
|---|---|---|---|
| a | 11.08 | -0.175 | 0.154 |
| b | -0.175 | 11.175 | -0.058 |
| c | 0.153 | -0.058 | 11.134 |

Fractional coordinates of the 64 atoms of Si and 1 atom of Na

| | a | b | c |
|---|---|---|---|
| Si | 0.359 | 0.924 | 0.793 |
| Si | 0.343 | 0.645 | 0.339 |
| Si | 0.295 | 0.722 | 0.846 |
| Si | 0.642 | 0.594 | 0.199 |
| Si | 0.457 | 0.489 | 0.252 |
| Si | 0.734 | 0.918 | 0.253 |
| Si | 0.549 | 0.246 | 0.513 |
| Si | 0.821 | 0.465 | 0.596 |
| Si | 0.540 | 0.724 | 0.069 |
| Si | 0.584 | 0.186 | 0.850 |
| Si | 0.739 | 0.728 | 0.347 |
| Si | 0.567 | 0.968 | 0.818 |
| Si | 0.280 | 0.087 | 0.105 |
| Si | 0.795 | 0.190 | 0.814 |
| Si | 0.828 | 0.918 | 0.059 |
| Si | 0.259 | 0.084 | 0.893 |
| Si | 0.437 | 0.598 | 0.935 |
| Si | 0.391 | 0.433 | 0.059 |
| Si | 0.676 | 0.831 | 0.941 |
| Si | 0.261 | 0.284 | 0.820 |
| Si | 0.343 | 0.934 | 0.578 |
| Si | 0.326 | 0.609 | 0.666 |
| Si | 0.141 | 0.985 | 0.536 |
| Si | 0.902 | 0.368 | 0.765 |
| Si | 0.895 | 0.164 | 0.487 |
| Si | 0.630 | 0.749 | 0.538 |
| Si | 0.058 | 0.021 | 0.894 |
| Si | 0.762 | 0.319 | 0.448 |
| Si | 0.553 | 0.287 | 0.039 |
| Si | 0.093 | 0.698 | 0.910 |
| Si | 0.006 | 0.940 | 0.704 |
| Si | 0.723 | 0.424 | 0.095 |
| Si | 0.192 | 0.373 | 0.011 |
| Si | 0.048 | 0.530 | 0.032 |
| Si | 0.462 | 0.296 | 0.717 |
| Si | 0.176 | 0.502 | 0.368 |
| Si | 0.041 | 0.861 | 0.031 |
| Si | 0.060 | 0.900 | 0.353 |
| Si | 0.528 | 0.972 | 0.282 |
| Si | 0.852 | 0.371 | 0.252 |
| Si | 0.836 | 0.051 | 0.655 |
| Si | 0.217 | 0.297 | 0.355 |
| Si | 0.106 | 0.330 | 0.686 |
| Si | 0.951 | 0.707 | 0.374 |
| Si | 0.899 | 0.037 | 0.319 |
| Si | 0.491 | 0.505 | 0.748 |
| Si | 0.517 | 0.036 | 0.491 |
| Si | 0.389 | 0.809 | 0.205 |
| Si | 0.675 | 0.595 | 0.678 |
| Si | 0.653 | 0.945 | 0.620 |
| Si | 0.431 | 0.295 | 0.345 |
| Si | 0.154 | 0.513 | 0.581 |
| Si | 0.147 | 0.235 | 0.165 |
| Si | 0.888 | 0.112 | 0.985 |
| Si | 0.195 | 0.902 | 0.177 |
| Si | 0.940 | 0.191 | 0.176 |
| Si | 1.005 | 0.537 | 0.250 |
| Si | 0.410 | 0.738 | 0.520 |
| Si | 0.974 | 0.618 | 0.568 |
| Si | 0.964 | 0.736 | 0.746 |
| Si | 0.469 | 0.152 | 0.187 |
| Si | 0.105 | 0.198 | 0.515 |
| Si | 0.780 | 0.673 | 0.842 |
| Na | 0.858 | 0.495 | 0.938 |

Mg

Lattice vectors (Å)

| | | | |
|---|---|---|---|
| a | 11.095 | -0.157 | 0.198 |
| b | -0.157 | 11.203 | -0.074 |
| c | 0.198 | -0.074 | 11.086 |

Fractional coordinates of the 64 atoms of Si and 1 atom of Mg

| | a | b | c |
|---|---|---|---|
| Si | 0.353 | 0.922 | 0.791 |
| Si | 0.333 | 0.647 | 0.336 |
| Si | 0.285 | 0.723 | 0.843 |
| Si | 0.636 | 0.596 | 0.198 |
| Si | 0.456 | 0.493 | 0.262 |
| Si | 0.733 | 0.921 | 0.251 |
| Si | 0.538 | 0.252 | 0.513 |
| Si | 0.815 | 0.464 | 0.594 |
| Si | 0.534 | 0.728 | 0.071 |
| Si | 0.580 | 0.184 | 0.850 |
| Si | 0.734 | 0.730 | 0.341 |
| Si | 0.561 | 0.966 | 0.816 |
| Si | 0.279 | 0.086 | 0.103 |
| Si | 0.789 | 0.188 | 0.811 |
| Si | 0.821 | 0.920 | 0.054 |
| Si | 0.255 | 0.082 | 0.891 |
| Si | 0.433 | 0.602 | 0.934 |
| Si | 0.390 | 0.439 | 0.064 |
| Si | 0.669 | 0.831 | 0.939 |
| Si | 0.257 | 0.278 | 0.813 |
| Si | 0.339 | 0.935 | 0.576 |
| Si | 0.318 | 0.608 | 0.663 |
| Si | 0.137 | 0.985 | 0.533 |
| Si | 0.895 | 0.367 | 0.762 |
| Si | 0.887 | 0.162 | 0.484 |
| Si | 0.620 | 0.746 | 0.531 |
| Si | 0.054 | 0.020 | 0.891 |
| Si | 0.751 | 0.318 | 0.451 |
| Si | 0.549 | 0.290 | 0.038 |
| Si | 0.080 | 0.698 | 0.904 |
| Si | 0.000 | 0.940 | 0.700 |
| Si | 0.716 | 0.425 | 0.093 |
| Si | 0.195 | 0.375 | 0.007 |
| Si | 0.040 | 0.526 | 0.028 |
| Si | 0.462 | 0.300 | 0.720 |
| Si | 0.169 | 0.500 | 0.367 |
| Si | 0.033 | 0.862 | 0.030 |
| Si | 0.058 | 0.898 | 0.349 |
| Si | 0.533 | 0.985 | 0.286 |
| Si | 0.841 | 0.368 | 0.252 |
| Si | 0.829 | 0.048 | 0.652 |
| Si | 0.211 | 0.296 | 0.352 |
| Si | 0.099 | 0.329 | 0.683 |
| Si | 0.945 | 0.707 | 0.371 |
| Si | 0.90 | 0.036 | 0.315 |
| Si | 0.484 | 0.508 | 0.743 |
| Si | 0.514 | 0.041 | 0.494 |
| Si | 0.394 | 0.822 | 0.215 |
| Si | 0.665 | 0.593 | 0.667 |
| Si | 0.647 | 0.941 | 0.618 |
| Si | 0.424 | 0.30 | 0.346 |
| Si | 0.147 | 0.512 | 0.581 |
| Si | 0.145 | 0.235 | 0.160 |
| Si | 0.883 | 0.112 | 0.981 |
| Si | 0.195 | 0.899 | 0.173 |
| Si | 0.937 | 0.192 | 0.172 |
| Si | 0.996 | 0.532 | 0.250 |
| Si | 0.402 | 0.738 | 0.518 |
| Si | 0.968 | 0.617 | 0.564 |
| Si | 0.952 | 0.737 | 0.741 |
| Si | 0.465 | 0.157 | 0.187 |
| Si | 0.097 | 0.196 | 0.511 |
| Si | 0.768 | 0.673 | 0.835 |
| Mg | 0.848 | 0.497 | 0.933 |

F. 2. Crystalline Si structures

*F. 2. a) Pure crystalline Si structure, a 64 atom supercell*

Lattice vectors (Å)

| | | | |
|---|---|---|---|
| a | 11.005 | -0.007 | -0.007 |
| b | -0.007 | 11.005 | -0.007 |
| c | -0.007 | -0.007 | 11.005 |

Fractional coordinates of the 64 atoms of Si

|    | a      | b      | c      |
|----|--------|--------|--------|
| Si | -0.002 | -0.002 | -0.002 |
| Si | 0.123  | 0.123  | 0.123  |
| Si | -0.002 | 0.248  | 0.248  |
| Si | 0.123  | 0.373  | 0.373  |
| Si | 0.248  | -0.002 | 0.248  |
| Si | 0.373  | 0.123  | 0.373  |
| Si | 0.248  | 0.248  | -0.002 |
| Si | 0.373  | 0.373  | 0.123  |
| Si | 0.498  | -0.002 | -0.002 |
| Si | 0.623  | 0.123  | 0.123  |
| Si | 0.498  | 0.248  | 0.248  |
| Si | 0.623  | 0.372  | 0.372  |
| Si | 0.748  | -0.003 | 0.248  |
| Si | 0.873  | 0.122  | 0.372  |
| Si | 0.748  | 0.248  | -0.003 |
| Si | 0.873  | 0.372  | 0.122  |
| Si | -0.002 | 0.498  | -0.002 |
| Si | 0.123  | 0.623  | 0.123  |
| Si | -0.003 | 0.748  | 0.248  |
| Si | 0.122  | 0.873  | 0.372  |
| Si | 0.248  | 0.498  | 0.248  |
| Si | 0.372  | 0.623  | 0.372  |
| Si | 0.248  | 0.748  | -0.003 |
| Si | 0.372  | 0.873  | 0.122  |
| Si | 0.498  | 0.498  | -0.003 |
| Si | 0.623  | 0.623  | 0.122  |
| Si | 0.497  | 0.747  | 0.247  |
| Si | 0.622  | 0.872  | 0.372  |
| Si | 0.747  | 0.497  | 0.247  |
| Si | 0.872  | 0.622  | 0.372  |
| Si | 0.747  | 0.747  | -0.003 |
| Si | 0.872  | 0.872  | 0.122  |
| Si | -0.002 | -0.002 | 0.498  |
| Si | 0.123  | 0.123  | 0.623  |
| Si | -0.003 | 0.248  | 0.748  |
| Si | 0.122  | 0.372  | 0.873  |
| Si | 0.248  | -0.003 | 0.748  |
| Si | 0.372  | 0.122  | 0.873  |
| Si | 0.248  | 0.248  | 0.498  |
| Si | 0.372  | 0.372  | 0.623  |
| Si | 0.498  | -0.003 | 0.498  |
| Si | 0.623  | 0.122  | 0.623  |
| Si | 0.497  | 0.247  | 0.747  |
| Si | 0.622  | 0.372  | 0.872  |
| Si | 0.747  | -0.003 | 0.747  |
| Si | 0.872  | 0.122  | 0.872  |
| Si | 0.747  | 0.247  | 0.497  |
| Si | 0.872 | 0.372 | 0.622 |
| Si | -0.003 | 0.498 | 0.498 |
| Si | 0.122 | 0.623 | 0.623 |
| Si | -0.003 | 0.747 | 0.747 |
| Si | 0.122 | 0.872 | 0.872 |
| Si | 0.247 | 0.497 | 0.747 |
| Si | 0.372 | 0.622 | 0.872 |
| Si | 0.247 | 0.747 | 0.497 |
| Si | 0.372 | 0.872 | 0.622 |
| Si | 0.497 | 0.497 | 0.497 |
| Si | 0.622 | 0.622 | 0.622 |
| Si | 0.497 | 0.747 | 0.747 |
| Si | 0.622 | 0.872 | 0.872 |
| Si | 0.747 | 0.497 | 0.747 |
| Si | 0.872 | 0.622 | 0.872 |
| Si | 0.747 | 0.747 | 0.497 |
| Si | 0.872 | 0.872 | 0.622 |

*F. 2. b) Lowest energy site structures*

Li

Lattice vectors (Å)

| | | | |
|---|---|---|---|
| a | 11.02 | 0.000 | 0.000 |
| b | 0.000 | 11.02 | 0.000 |
| c | 0.000 | 0.000 | 11.02 |

Fractional coordinates of the 64 atoms of Si and 1 atom of Li

| | a | b | c |
|---|---|---|---|
| Si | 0.374 | 0.125 | 0.374 |
| Si | 0.370 | 0.370 | 0.630 |
| Si | 0.626 | 0.125 | 0.626 |
| Si | 0.630 | 0.370 | 0.370 |
| Si | 0.500 | 0.248 | 0.500 |
| Si | 0.752 | 0.000 | 0.500 |
| Si | 0.500 | 0.500 | 0.752 |
| Si | 0.750 | 0.250 | 0.750 |
| Si | 0.876 | 0.124 | 0.376 |
| Si | 0.875 | 0.374 | 0.626 |
| Si | 0.124 | 0.124 | 0.624 |
| Si | 0.125 | 0.374 | 0.374 |
| Si | 0.000 | 0.250 | 0.500 |
| Si | 0.248 | 0.500 | 0.500 |
| Si | 0.000 | 0.500 | 0.750 |
| Si | 0.247 | 0.247 | 0.753 |
| Si | 0.370 | 0.630 | 0.370 |
| Si | 0.374 | 0.875 | 0.626 |
| Si | 0.630 | 0.630 | 0.630 |
| Si | 0.626 | 0.875 | 0.374 |
| Si | 0.500 | 0.752 | 0.500 |
| Si | 0.750 | 0.000 | 0.500 |
| Si | 0.500 | 0.000 | 0.750 |
| Si | 0.753 | 0.753 | 0.753 |
| Si | 0.376 | 0.124 | 0.876 |
| Si | 0.374 | 0.374 | 0.125 |
| Si | 0.624 | 0.124 | 0.124 |
| Si | 0.626 | 0.374 | 0.875 |
| Si | 0.500 | 0.250 | 0.000 |
| Si | 0.750 | 0.500 | 0.000 |
| Si | 0.500 | 0.500 | 0.248 |
| Si | 0.753 | 0.247 | 0.247 |
| Si | 0.875 | 0.125 | 0.875 |
| Si | 0.876 | 0.376 | 0.124 |
| Si | 0.125 | 0.125 | 0.125 |
| Si | 0.124 | 0.376 | 0.876 |
| Si | 0.000 | 0.250 | 0.000 |
| Si | 0.250 | 0.500 | 0.000 |
| Si | 0.000 | 0.500 | 0.250 |
| Si | 0.250 | 0.250 | 0.250 |
| Si | 0.374 | 0.626 | 0.875 |
| Si | 0.376 | 0.876 | 0.124 |
| Si | 0.626 | 0.626 | 0.125 |
| Si | 0.624 | 0.876 | 0.876 |
| Si | 0.500 | 0.750 | 0.000 |
| Si | 0.750 | 0.000 | 0.000 |
| Si | 0.500 | 0.000 | 0.250 |
| Si | 0.750 | 0.750 | 0.250 |
| Si | 0.875 | 0.626 | 0.374 |
| Si | 0.876 | 0.876 | 0.624 |
| Si | 0.125 | 0.626 | 0.626 |
| Si | 0.124 | 0.876 | 0.376 |
| Si | 0.000 | 0.750 | 0.500 |
| Si | 0.250 | 0.000 | 0.500 |
| Si | 0.000 | 0.000 | 0.750 |
| Si | 0.250 | 0.750 | 0.750 |
| Si | 0.876 | 0.624 | 0.876 |
| Si | 0.875 | 0.875 | 0.125 |
| Si | 0.124 | 0.624 | 0.124 |
| Si | 0.125 | 0.875 | 0.875 |
| Si | 0.000 | 0.750 | 0.000 |
| Si | 0.250 | 0.000 | 0.000 |
| Si | 0.000 | 0.000 | 0.250 |
| Si | 0.247 | 0.753 | 0.247 |
| Li | 0.500 | 0.500 | 0.500 |

Na

Lattice vectors (Å)

| | | | |
|---|---|---|---|
| a | 11.039 | 0.000 | 0.000 |
| b | 0.000 | 11.039 | 0.000 |
| c | 0.000 | 0.000 | 11.039 |

Fractional coordinates of the 64 atoms of Si and 1 atom of Na

| | a | b | c |
|---|---|---|---|
| Si | 0.374 | 0.124 | 0.374 |
| Si | 0.368 | 0.368 | 0.632 |
| Si | 0.626 | 0.124 | 0.626 |
| Si | 0.632 | 0.368 | 0.368 |
| Si | 0.500 | 0.246 | 0.500 |
| Si | 0.754 | 0.500 | 0.500 |
| Si | 0.500 | 0.500 | 0.754 |
| Si | 0.750 | 0.250 | 0.750 |
| Si | 0.876 | 0.124 | 0.376 |
| Si | 0.876 | 0.374 | 0.626 |
| Si | 0.124 | 0.124 | 0.624 |
| Si | 0.124 | 0.374 | 0.374 |
| Si | 0.000 | 0.250 | 0.500 |
| Si | 0.246 | 0.500 | 0.500 |
| Si | 0.000 | 0.500 | 0.750 |
| Si | 0.245 | 0.245 | 0.755 |
| Si | 0.368 | 0.632 | 0.368 |
| Si | 0.374 | 0.876 | 0.626 |
| Si | 0.632 | 0.632 | 0.632 |
| Si | 0.626 | 0.876 | 0.374 |
| Si | 0.500 | 0.754 | 0.500 |
| Si | 0.750 | 0.000 | 0.500 |
| Si | 0.500 | 0.000 | 0.750 |
| Si | 0.755 | 0.755 | 0.755 |
| Si | 0.376 | 0.124 | 0.876 |
| Si | 0.374 | 0.374 | 0.124 |
| Si | 0.624 | 0.124 | 0.124 |
| Si | 0.626 | 0.374 | 0.876 |
| Si | 0.500 | 0.250 | 0.000 |
| Si | 0.750 | 0.500 | 0.000 |
| Si | 0.500 | 0.500 | 0.246 |
| Si | 0.755 | 0.245 | 0.245 |
| Si | 0.875 | 0.125 | 0.875 |
| Si | 0.876 | 0.376 | 0.124 |
| Si | 0.125 | 0.125 | 0.125 |
| Si | 0.124 | 0.376 | 0.876 |
| Si | 0.000 | 0.250 | 0.000 |
| Si | 0.250 | 0.500 | 0.000 |
| Si | 0.000 | 0.500 | 0.250 |
| Si | 0.250 | 0.250 | 0.250 |
| Si | 0.374 | 0.626 | 0.876 |
| Si | 0.376 | 0.876 | 0.124 |
| Si | 0.626 | 0.626 | 0.124 |
| Si | 0.624 | 0.876 | 0.876 |
| Si | 0.500 | 0.750 | 0.000 |
| Si | 0.750 | 0.000 | 0.000 |
| Si | 0.500 | 0.000 | 0.250 |
| Si | 0.750 | 0.750 | 0.250 |
| Si | 0.876 | 0.626 | 0.374 |
| Si | 0.876 | 0.876 | 0.624 |
| Si | 0.124 | 0.626 | 0.626 |
| Si | 0.124 | 0.876 | 0.376 |
| Si | 0.000 | 0.750 | 0.500 |
| Si | 0.250 | 0.000 | 0.500 |
| Si | 0.000 | 0.000 | 0.750 |
| Si | 0.250 | 0.750 | 0.750 |
| Si | 0.876 | 0.624 | 0.876 |
| Si | 0.875 | 0.875 | 0.125 |
| Si | 0.124 | 0.624 | 0.124 |
| Si | 0.125 | 0.875 | 0.875 |
| Si | 0.000 | 0.750 | 0.000 |
| Si | 0.250 | 0.000 | 0.000 |
| Si | 0.000 | 0.000 | 0.250 |
| Si | 0.245 | 0.755 | 0.245 |
| Na | 0.500 | 0.500 | 0.500 |

Mg

Lattice vectors (Å)

| | | | |
|---|---|---|---|
| a | 11.034 | 0.000 | 0.000 |
| b | 0.000 | 11.034 | 0.000 |
| c | 0.000 | 0.000 | 11.034 |

Fractional coordinates of the 64 atoms of Si and 1 atom of Mg

| | a | b | c |
|---|---|---|---|
| Si | 0.374 | 0.124 | 0.374 |
| Si | 0.368 | 0.368 | 0.632 |
| Si | 0.626 | 0.124 | 0.626 |
| Si | 0.632 | 0.368 | 0.368 |
| Si | 0.500 | 0.246 | 0.500 |
| Si | 0.754 | 0.500 | 0.500 |
| Si | 0.500 | 0.500 | 0.754 |
| Si | 0.750 | 0.250 | 0.750 |
| Si | 0.876 | 0.124 | 0.376 |
| Si | 0.876 | 0.374 | 0.626 |
| Si | 0.124 | 0.124 | 0.624 |
| Si | 0.124 | 0.374 | 0.374 |
| Si | 0.000 | 0.250 | 0.500 |
| Si | 0.246 | 0.500 | 0.500 |
| Si | 0.000 | 0.500 | 0.750 |
| Si | 0.244 | 0.244 | 0.756 |
| Si | 0.368 | 0.632 | 0.368 |
| Si | 0.374 | 0.876 | 0.626 |
| Si | 0.632 | 0.632 | 0.632 |
| Si | 0.626 | 0.876 | 0.374 |
| Si | 0.500 | 0.754 | 0.500 |
| Si | 0.750 | 0.000 | 0.500 |
| Si | 0.500 | 0.000 | 0.750 |
| Si | 0.756 | 0.756 | 0.756 |
| Si | 0.376 | 0.124 | 0.876 |
| Si | 0.374 | 0.374 | 0.124 |
| Si | 0.624 | 0.124 | 0.124 |
| Si | 0.626 | 0.374 | 0.876 |
| Si | 0.500 | 0.250 | 0.000 |
| Si | 0.750 | 0.500 | 0.000 |
| Si | 0.500 | 0.500 | 0.246 |
| Si | 0.756 | 0.244 | 0.244 |
| Si | 0.875 | 0.125 | 0.875 |
| Si | 0.876 | 0.376 | 0.124 |
| Si | 0.125 | 0.125 | 0.125 |
| Si | 0.124 | 0.376 | 0.876 |
| Si | 0.000 | 0.250 | 0.000 |
| Si | 0.250 | 0.500 | 0.000 |
| Si | 0.000 | 0.500 | 0.250 |
| Si | 0.250 | 0.250 | 0.250 |
| Si | 0.374 | 0.626 | 0.876 |
| Si | 0.376 | 0.876 | 0.124 |
| Si | 0.626 | 0.626 | 0.124 |
| Si | 0.624 | 0.876 | 0.876 |
| Si | 0.500 | 0.750 | 0.000 |
| Si | 0.750 | 0.000 | 0.000 |
| Si | 0.500 | 0.000 | 0.250 |
| Si | 0.750 | 0.750 | 0.250 |
| Si | 0.876 | 0.626 | 0.374 |
| Si | 0.876 | 0.876 | 0.624 |
| Si | 0.124 | 0.626 | 0.626 |
| Si | 0.124 | 0.876 | 0.376 |
| Si | 0.000 | 0.750 | 0.500 |
| Si | 0.250 | 0.000 | 0.500 |
| Si | 0.000 | 0.000 | 0.750 |
| Si | 0.250 | 0.750 | 0.750 |
| Si | 0.876 | 0.624 | 0.876 |
| Si | 0.875 | 0.875 | 0.125 |
| Si | 0.124 | 0.624 | 0.124 |
| Si | 0.125 | 0.875 | 0.875 |
| Si | 0.000 | 0.750 | 0.000 |
| Si | 0.250 | 0.000 | 0.000 |
| Si | 0.000 | 0.000 | 0.250 |
| Si | 0.244 | 0.756 | 0.244 |
| Mg | 0.500 | 0.500 | 0.500 |

# G. Lattice parameters and cohesive energies* of metals

| | Lattice parameters (Å) | | Cohesive energy (eV) | | |
|---|---|---|---|---|---|
| | SIESTA | exp. | SIESTA | Other DFT | exp. |
| Li | a = 3.68 | a = 3.51 | 1.80 | 1.20-2.01 [4] | 1.69 |
| Na | a = 4.24 | a = 4.29 | 1.39 | 0.73-1.21 [5] | 1.13 |
| Mg | a = 3.25<br>c = 5.26 | a = 3.21<br>c = 5.21 | 1.59 | 1.42-1.78 [6] | 1.51 |

*electronic only